 \documentclass[final,5p,times,twocolumn]{elsarticle}

\usepackage{amssymb}
\usepackage{lipsum}

\usepackage{graphicx}
\usepackage{bm}
\usepackage[utf8]{inputenc}
\usepackage{amsmath}
\usepackage{relsize}
\usepackage{multirow}
\usepackage{xfrac}
\usepackage[colorlinks=true
]{hyperref}
\usepackage{orcidlink}

\newcommand*{\bi}{\bibitem}
\newcommand*{\ea}{\textit{et al.}}
\newcommand*{\eg}{\textit{e.g.}}
\newcommand*{\ra}{\rightarrow}
\newcommand*{\dspdsm}{D_s^+D_s^-}

\newcommand*{\rf}[1]{(\ref{#1})}
\newcommand*{\be}{\begin{equation}}
\newcommand*{\ee}{\end{equation}}

\newcommand*{\die}{e^+e^-}
\newcommand*{\jj}{\mathrm i}
\newcommand*{\cndf}{\chi^2/\mathrm{DOF}}
\newcommand*{\ndf}{\mathrm{DOF}}
\newcommand*{\minuit}{\texttt{MINUIT}~}

\newcommand*{\dek}[1]{\times10^{#1}}
\newcommand{\IM}{\mathrm{Im}}
\newcommand{\p}{$p$-value }
\newcommand*{\bea}{\begin{eqnarray*}}
\newcommand*{\eea}{\end{eqnarray*}}
\newcommand*{\dd}{\die\to D\bar D}

\journal{Physics Letters B}

\begin{document}

\begin{frontmatter}



\title{A subthreshold pole in electron‐positron annihilation into
 \texorpdfstring{\bm{${D_s^+D_s^-}$}}{Ds+ Ds-} final state}


\author[first]{Peter Lichard\,\orcidlink{0000-0003-1581-8545}\,}
\affiliation[first]{organization={Institute of Physics and Research Centre for Computational Physics and Data Processing, Silesian University in Opava, 746 01 Opava},
            country={Czech Republic}}
\makeatletter \@eadauthor={Peter Lichard} \makeatother
\ead{peter.lichard@physics.slu.cz}

\author[second]{Josef Jur\'a\v{n}\,\orcidlink{0000-0002-0444-9716}\,\corref{cor2}}
\affiliation[second]{organization={Institute of Physics, Silesian University in Opava, 746 01 Opava},
            country={Czech Republic}}
\makeatletter \@eadauthor={Josef Jur\'a\v{n}} \makeatother
\ead{josef.juran@physics.slu.cz}
\cortext[cor2]{Corresponding author.}

\begin{abstract}
Using precise data from the BESIII collaboration, we obtained compelling
evidence of a subthreshold pole in the $e^+ e^-\rightarrow  D_s^+ D_s^-$
process with a mass of (3896$^{+13}_{-22}$)~MeV. Its interpretation in terms
of the existing physical states is open.
\end{abstract}



\begin{keyword}
Electron-positron annihilation to hadrons \sep Subtreshold poles \sep Resonances


\end{keyword}

\end{frontmatter}




\section{Introduction}
\label{introduction}

This paper is the next step in our effort to uncover the role of the
subthreshold poles (SP, in what follows) in the electron-positron annihilation
into hadrons \cite{kaonium,pionium,dpdm}.

The playground in our consideration is the complex $s$ plane in which
the $s$ variable acquires the values of the cms energy squared on the real
axis from the reaction threshold to infinity. We accept as a working
hypothesis that the amplitude of the $\die$ annihilation into a hadronic
final state as a function of $s$ has the same analytic properties as
the amplitudes of hadronic reactions, discussed, \eg, in
\cite{eden,pilkuhn,mizera}.
Namely, that the annihilation amplitude, when continued to the whole
complex $s$ plane, becomes an analytic function
with a cut along the above-mentioned part of the real axis (called the physical
cut). The imaginary part of the amplitude exhibits a discontinuity across the
physical cut, $\IM\,F(s+\jj\epsilon)\ne \IM\,F(s-\jj\epsilon)$. 

The resonances, which reveal themselves as pronounced structures in the
excitation function, are represented by pairs of complex-conjugated poles on
higher Riemann sheets that are accessible through the physical cut. Their
contributions to the amplitude on the upper branch of the physical cut are
usually described by various phenomenological formulas.

Also, the poles on the first (physical) Riemann sheet may exist, but only
on the real axis below the reaction threshold. These SPs correspond to states
that are stable against the decay to the considered final state. They are
not reachable in the experiment and are
not usually considered when fitting the experimental data. However, they may
influence the amplitude in the physical region due to strong restrictions
implied by analytic properties of amplitude, which are the consequence
of causality and unitarity.

We aim to discover the SPs by fitting the experimental data on the $\die$
annihilation cross section. In this paper, we concentrate on the annihilation
into the $\dspdsm$ final state. 

\section{The explored data and the model}
\label{Data-Model}

The BESIII collaboration working at the spectrometer situated at
the electron-positron collider BEPCII, based at the IHEP laboratory
in Beijing, China, have recently published \cite{besiii2024b} precise data
on $\die$ annihilation to the $\dspdsm$ final states. In the Supplemental
Material \cite{supplemental}, they provide the Born cross sections of the
$\die\ra\dspdsm$ at 138 energies from the threshold at 3.938~GeV to 4.951~GeV
with both statistical and systematic errors.

To analyze the BESIII data, we will use the cross-section formula derived in
\cite{dpdm}. Replacing the coefficients $Q_k$ in it by $C_k=\pi\alpha^2Q_k/3$,
we cast it to a more familiar form
\be
\label{sigma}
\sigma(s)=\frac{\beta^3}{s}
\left|\sum_{k=1}^n
\frac{\sqrt{C_k}\,e^{\jj\delta_k}}{s-M_k^2+\jj M_k\Gamma_k}
\right|^2\!.
\ee
Here, $M_k$ and $\Gamma_k$ are the mass and width of the $k$th
intermediate vector meson coupled to the photon, respectively, and
\be
\label{beta}
\beta=\sqrt{1-\frac{4m_{D_s}^2}{s}}
\ee
is the speed of an outgoing meson. The formula \rf{sigma} allows both the
SP and resonances in the intermediate state. For the former,
$\Gamma_k=0$, for the latter, the coefficient $C_k$ can be related to measurable
quantities by
\be
\label{ck}
C_k=\frac{12\pi M_k^2}{\beta_k^3}\Gamma_k{\cal B}_{V_k\ra D_s^+ D_s^-}
\Gamma_{V_k\ra e^+e^-},
\ee
where $\beta_k$ is given by Eq. \rf{beta} with $s=M_k^2$.

We will treat $M_k$s, $\Gamma_k$s, $C_k$s, and $\delta_k$s as
free parameters (except $\delta_1$, which is kept at 0). We will
determine their mean values and dispersions by fitting the formula \rf{sigma}
to experimental cross sections using the standard $\chi^2$ criterion.
We use the program \minuit by F. James and M. Roos \cite{minuit}.
To fit the data, we merge their statistical and systematic errors
quadratically.  For the mass of the $D_s^\pm$, we will use the value
$m_{D_s}=(1969.0\pm1.4)$~MeV, listed in the PDG 2026 Tables
\cite{pdg2026} as ``our average''.

\section{The Results}
\label{results}

To fit our model to the BESIII data, we began with the formula \rf{sigma}
featuring a single resonance. Then, we used the obtained parameters as input
in the next step, where we added the second resonance. Repeating the
procedure, we arrived at a three-resonance (3R) fit characterized by
$\cndf=230.5/127$ and \p of 5$\dek{-8}$. Assuming the fourth
resonance (4R) improved the fit to $\cndf=166.2/123$ (\p = 0.6\%).

The parameters of the fourth 'resonance' emerged as a significant surprise:
$M=(3891\pm13)$~MeV and $\Gamma=(0\pm110)$~MeV. The fact that $\Gamma$ is
compatible with zero suggests that the corresponding pole lies on the real
$s$ axis at $B=(47\pm13)$~MeV below the $\dspdsm$ threshold. This unexpected
result indicates that to enhance the fit quality, an SP is required, not
a resonance. 
The statistical significance of the SP, which we
estimated by utilizing the changes in likelihood values
$\delta(-2\ln{\cal L})=64.3$ and in the number of degrees $\delta(\ndf)=4$,
is a striking $7.2\sigma$.


Following this lead, we adjusted the fit by placing the SP in the first
position ($k=1$) with $\Gamma_1=\delta_1=0$ and increased the number of resonances to four (SP+4R fit). We obtained $\cndf=124.5/120$, with
\p of 37\%.
We show the parameters of the fit in Table \ref{tab:p4r}.
\begin{table*}
\begin{tabular}{l c c c c c}
\hline
\multicolumn{6}{c}{$\cndf=124.5/120$~~~~~~~\p = 37\%}\\
\hline
$k$ & 1$\equiv$SP & 2 & 3 & 4 & 5  \\
\hline
$M_k$ (MeV) &~3896$\pm$12~&~4026.4$\pm$1.6~&~4212.9$\pm$4.7~&~4237.2$\pm$9.5~&
~4417.9$\pm$5.8 \\
$\Gamma_k$ (MeV)& 0(f) & 36.7$\pm$3.2 & 48$\pm$10 & 115$\pm$24 & 44$\pm$15
\\
$C_k\times10^6$ (GeV$^4$) & 363$\pm$80 & 38.1$\pm$8.0 & 4.5$\pm$5.4 & 51$\pm$21
 & 0.081(58) \\
$\delta_k$ (rad) & 0(f)& 2.96(20) & 0.63(52) & $-1.45(22)$ & $-0.78(40)$ \\
\hline
\end{tabular}
\caption{Parameters of the SP+4R fit to the BESIII
$\die\ra\dspdsm$ data \cite{besiii2024b} using the formula \rf{sigma}. Symbol (f) marks values that we kept fixed.
}
\label{tab:p4r}
\end{table*}
As the SP mass is the most important outcome of our study, we performed a
more detailed study of its errors using the $\chi^2_\mathrm{min}$+1
criterion. We end up with $M=(3896^{+13}_{-22})$~MeV.

\begin{figure*}
	\centering
	\includegraphics[width=1\textwidth, angle=0]{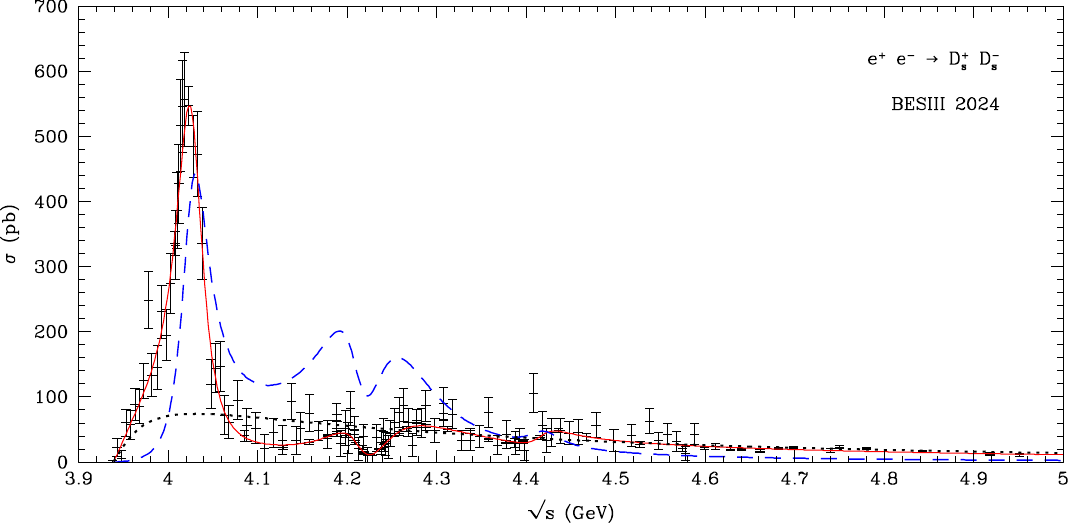}
	\caption{The $\die\ra\dspdsm$ cross section calculated
assuming an SP and four resonances (full, red online) and its comparison to
data. Also, the individual contributions from the resonance part (dashed,
blue online) and SP (dotted) are shown.}
	\label{fig:p4r}%
\end{figure*}

In Fig. \ref {fig:p4r}, the excitation function representing the SP+4R fit
is displayed in comparison with the data. In
addition, the resonance part (calculated from Eq. \rf{sigma} with
$C_1$ set to 0) and the subthreshold part ($C_2=\ldots=C_5=0$) are shown.
The important role of the SP is confirmed. The interference between the SP and resonance parts is essential in obtaining the final shape of the excitation function.

In an effort to find a better solution (in a $\chi^2$ sense), we repeated
the minimization procedure many times with randomly chosen starting
parameters. We found none. But we found several solutions with the same
$\chi^2$, $M_k$s, and $\Gamma_k$s\footnote{\,The differences between the
central values of the corresponding parameters across different solutions are much smaller than their errors.}, substantially differing in values of
$C_k$s and $\delta_k$s. This phenomenon of multiple solutions is explained
in the literature \cite{bukin, zhu,han, bai}. The set of nine parameters
$C_k$ and $\delta_k$ presented in Table \ref{tab:p4r} is just a randomly
chosen one among the many possible sets, which, without additional
information, are equally good. To choose a preferred one, we would need to
know the product ${\cal B}_{V_k\ra D_s^+ D_s^-}\Gamma_{V_k\ra e^+e^-}$
[see Eq. \rf{ck}] for at least one resonance.

The phenomenon of multiple solutions does not influence our reasoning
concerning the presence of an SP, which is based on the $\chi^2$ and
the values of masses and widths.

\begin{figure*}
	\centering
	\includegraphics[width=1\textwidth, angle=0]{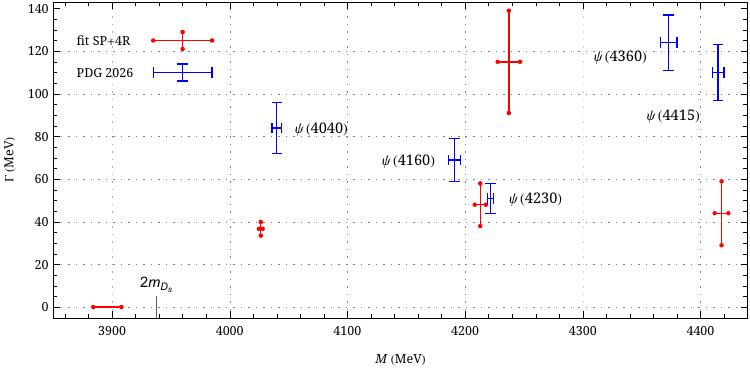}
	\caption{Parameters of resonances in the $\Gamma-M$ plane
obtained from the SP+4R fit (red) along the parameters of resonances from
PDG 2026 \cite{pdg2026} (blue).}
	\label{fig:ResGM}%
\end{figure*}

The main parameters of the resonances (masses and decay widths) obtained
in SP+4R fit are compared with resonances' parameters from PDG 2026
\cite{pdg2026} in Fig. \ref{fig:ResGM}.

Comparing our results with the PDG set of resonances, we are finding a big
discrepancy with PDG's $\psi(4040)$. The PDG based its values on the
results of a single experiment \cite{bes2008} done in 2008.
On the contrary, we explore recent, very precise data
by the BESIII Collaboration \cite{besiii2024b}. Our analysis is very stable
and gives consistent results under different assumptions on the fitting
function, as it is demonstrated in Table \ref{tab:psi4040}.

\begin{table}
\begin{tabular}{l c c c c}
 \hline
Fit~~~~&~~$\cndf$~&~\p~~~&~$M$~(MeV)~~&~~$\Gamma$~(MeV) \\
 \hline
2R    & 663.4/131 & 0           & $4017.1\pm1.9$ & $37.4\pm4.7$\\
3R    & 230.5/127 & 5$\dek{-8}$ & $4020.5\pm2.2$ & $49.3\pm6.5$\\
4R    & 166.2/123 & 0.6\%       & $4026.6\pm2.0$ & $36.3\pm4.0$\\
SP+4R & 124.5/120 & 37\%        & $4026.4\pm1.6$ & $36.7\pm3.2$\\
 \hline
\end{tabular}
\caption{The $\psi(4040)$ parameters obtained in various
fits to the BESIII data \cite{besiii2024b}. Errors are rescaled by
$\sqrt{\cndf}$.
}
\label{tab:psi4040}
\end{table}

Let us also note that our value of the $\psi(4040)$ mass is in the perfect
agreement with $M=(4026.6\pm1.9)$~MeV obtained in Ref. \cite{dpdm},
where the $\dd$ data by BES \cite{bes2008a} and BESIII \cite{besiii2024a}
collaborations were analysed. The width is a little higher here than
$\Gamma=(29.4\pm3.1)$~MeV in \cite{dpdm}. 
Similar values were obtained also in \cite{satoshi},
namely $M=(4029.2\pm0.4)$~MeV and $\Gamma=(26.3\pm1.0)$~MeV.

\section{Discussion}
\label{Discussion}

We have found strong evidence of an SP in the $\die\ra\dspdsm$
process with mass $M=(3896^{+13}_{-22})$~MeV (taking the value from the SP+4R fit,
which has an acceptable \p of 37\%). As our approach is purely phenomenological,
we cannot say anything about its origin. Below, we discuss a few
possibilities.

Given that the most recent compilation \cite{pdgxsect} of the $\die$
annihilation cross section does not show any vector charmonium state in the
corresponding CMS energy region, the identification of our SP with a
charmonium is excluded.

A possibility is that the SP we have found represents a bound
state in the $\dspdsm$ system with binding energy
$B=2m_{D^+_s}-M=(42^{+22}_{-13})$~MeV. Its quantum
numbers, dictated by the appearance in the $\die$ annihilation, are
$I^G(J^{PC})=0^-(1^{--})$. They exclude it as a ground state of the
$\dspdsm$ molecule, but suggest it as an excited state with orbital quantum
number $L=1$. The ground state should have a smaller mass and quantum numbers
$I^G(J^{PC})=0^+(0^{++})$.

Consulting the PDG Tables \cite{pdg2026}, we found a suitable candidate with
the right mass and quantum numbers in $\chi_{c0}(3860)$, which was observed by
the Belle Collaboration \cite{belle2017}. However, the LHCb Collaboration
\cite{lhcb2020} found no evidence of it, and at present, it is omitted from
the PDG Summary Tables \cite{pdg2026}.

In 2015, the $\chi_{c0}(3915)$ [aka $X(3915)$] was suggested to be an $S$-wave molecular bound state of the $D_s\bar{D}_s$ meson pair \cite{misha} by
analyzing its decay modes\footnote{\,Extensive citations to works on this topic can be found in \cite{wang,qi}.}. The $\chi_{c0}(3915)$ mass is $3922.1\pm1.8$~MeV, and the mass of an excited state must be higher than that.
This excludes our SP, whose mass of is lower by $26^{+22}_{-13}$~MeV.

The fact that the SP mass [$(3896^{+13}_{-22})$~MeV] is close to that of the
$G$(3900) resonance established at $(3872.5\pm14.2\pm3.0$)~MeV by BESIII
Collaboration \cite{besiii2024a} and to $(3898.4\pm0.9)$~MeV, recently found
in the global coupled-channel analysis \cite{satoshi} could mean that the
$G$(3900), which manifests itself as a resonance in the $\dd$
\cite{besiii2024a} and  $\die\ra D\bar D^\ast$/$\bar DD^\ast$ \cite{prl133}
processes, acts in the $\die\ra\dspdsm$ one as an SP. This
interpretation is in disagreement with the analysis \cite{dynamically}
that suggested the $G(3900)$ to be a P-wave dynamically generated state. In
fact, to become an SP, an entity must preexist as a state in the Fock space.

\section{Conclusions}
\label{Conclusions}

Our results provide evidence of the presence of an SP in the amplitude of the
$\die\ra\dspdsm$ process, which significantly influences the excitation curve.
Its mass came from the 4R fit as $M=(3891\pm13)$~MeV. The comparison of
the 3R and 4R fits gives the SP statistical significance of $7.2\sigma$. 

Adding another resonance has improved the quality of the fit (\p increased to 37\%). This SP+4R fit yields the mass of the SP pole $M=(3896^{+13}_{-22})$~MeV.

To conclude: We have shown with a statistical significance of 7.2$\sigma$
that a simple pole situated on the real $s$ axis below the $\dspdsm$
threshold exists.
But its identification with an existing physical state remains unclear.

\section*{Declaration of competing interest}
The authors declare that they have no known competing financial
interests or personal relationships that could have appeared to influence the work reported in this paper.

\section*{Acknowledgements}
This research did not receive any specific grant from funding agencies in the public, commercial, or not-for-profit sectors.

One of the authors (J.J.) thanks Satoshi X. Nakamura for enlightening
correspondence and Stanislav Hled\'{\i}k and Petr Se\v{d}a for discussions
about statistical issues.

\section*{Data availability}
The data analyzed in this article are openly available.





\end{document}